\documentstyle[12pt,world_sci]{article}
\input epsf
\begin{document}
\vsize=21.0truecm
\hsize=16.0truecm
\hspace{40cm} a.

\vspace{2truecm}

\noindent {{\large  Probing the Structure of Nucleons in Electromagnetic Interactions}}\\

\vspace{0.2truecm}\noindent
\noindent Volker D. Burkert\\


\noindent Jefferson Laboratory, 12000 Jefferson Avenue, Newport News, Virginia 23606, USA



\vspace{0.3truecm}

{I discuss open problems in nucleon structure studies using electromagnetic 
 probes. The focus is on the regime of strong interaction QCD. Significant progress 
 in our understanding of the nucleon structure in the region of strong QCD 
 may be expected in the first decade of the new millenium due to major 
 experimental and theoretical efforts currently underway in this field.}
   

\vspace{0.5truecm}

\section{INTRODUCTION}
                                                       
Electron scattering may be characterized according to 
distance 
and time scales (or momentum and energy transfer). 
At large distances mesons and nucleons are the relevant degrees of freedom. We can study 
peripheral properties of the nucleon near threshold, or meson exchange processes
at high energies.
 At short distances and short time scales, the coupling involves elementary quarks and 
gluon fields, governed  by pQCD, and we can map out parton distributions in the nucleon.  
At intermediate distances, quarks and gluons are 
relevant, however,  confinement is important, and they appear as constituent 
quarks and glue. We can study interactions between these constituents via their excitation spectra and 
wave functions. This is the region I will be focussing on. 
These regions obviously overlap, and the hope is that hadron structures may eventually
be described in a unified approach based on fundamental theory.
Because the electro-magnetic and -weak probes are well understood, they are best suited to
provide the data for such an endeavor. 
\vspace{0.5cm}

\subsection{ \bf Open problems in nucleon structure at intermediate distances}

QCD has not been solved for processes at intermediate distance scales and, 
therefore, the internal structure of nucleons is generally poorly known in this regime.
On the other hand, theorists are not challenged due to the lack 
of high quality data in many areas. The following are areas where the lack of high quality
data is most noticeable:

\begin{itemize}

\item{The electric form factors of the nucleon are poorly known, especially for the 
neutron, but also for the proton. This means that the charge distribution of the 
most common form of matter in the universe is virtually unknown.} 

\item{What role do strange quarks play in the wave function of ordinary 
matter?}

\item{The nucleon spin structure has been explored for more than two decades at high 
energies. The transition from the deep inelastic regime to the confinement 
regime has not been explored at all.}  

\item{To understand the ground state nucleon we need to understand
the full excitation spectrum as well as the continuum. 
Few transitions to excited states have been studied well, and many
states are missing from the spectrum as predicted by our most accepted models.} 

\item {The role of the glue in the baryon excitation spectrum is completely unknown, 
although gluonic excitations of the nucleon are likely produced copiously \cite{isgur}.}  
 
\item {The long-known connection between the deep inelastic regime and the regime of 
confinement (quark-hadron duality) \cite{blogil} remained virtually unexplored for decades.}

\end{itemize}

{\sl Carrying out an experimental program that will address these questions 
has become feasible due the 
availability of CW electron accelerator, modern detector instrumentation with 
high speed data transfer techniques, and the routine availability of spin
polarization. }

The main contributor to this field is now the CEBAF accelerator at Jefferson Lab 
in Newport News, Virginia, USA.
A maximum energy of 5.6 GeV is currently available, and the three
 experimental halls can receive polarized beam simultaneously, 
with different or the same beam energies.

\section{ELASTIC SCATTERING}

\subsection{\bf Electromagnetic form factors} 

This process probes the charge and current distribution in the nucleon 
in terms of the 
electric ($G^{\gamma}_E$)and magnetic ($G^{\gamma}_M$) form factors. 
Early experiments from Bonn, DESY, and CEA showed a violation of the so-called 
"scaling law", which 
may be interpreted that the spatial distribution  of charge and magnetization
are not the same, and the 
corresponding form factors have different $Q^2$ dependencies. The data showed a 
downward trend for the ratio $R^{\gamma}_{EM} = G^{\gamma}_E/G^{\gamma}_M$ as a function of $Q^2$.  
Adding the older and newer SLAC data sets confuses the picture greatly (Figure 1). 
Part of the  data 
are incompatible 
with the other data sets. They also do not show the same general trend as 
the other data sets. 
Reliable data were urgently needed to clarify the experimental situation and to constrain 
theoretical developments. 
 
The best way to get reliable data at high $Q^2$ is via double polarization experiments, 
and the first experiments of this type have now produced results. 
Since the ratio $R^{\gamma}_{EM}$ is accessed directly in the double polarization asymmetry 
$$A_{\vec e\vec p} = {k_1{R^{\gamma}_{EM}} \over {k_2(R^{\gamma}_{EM})^2 + k_3}}$$
this experiment has much lower systematic uncertainties than previous 
experiments at high $Q^2$ (Figure 1). They confirm the trend of the early data, 
improve the accuracy at high $Q^2$ significantly, and extend the $Q^2$ range. 
The data illustrate the power of utilizing polarization in electromagnetic 
interactions.

\begin{figure}[htbp]
\begin{minipage}{0.495\textwidth}
\epsfysize=6.5cm
\epsfbox{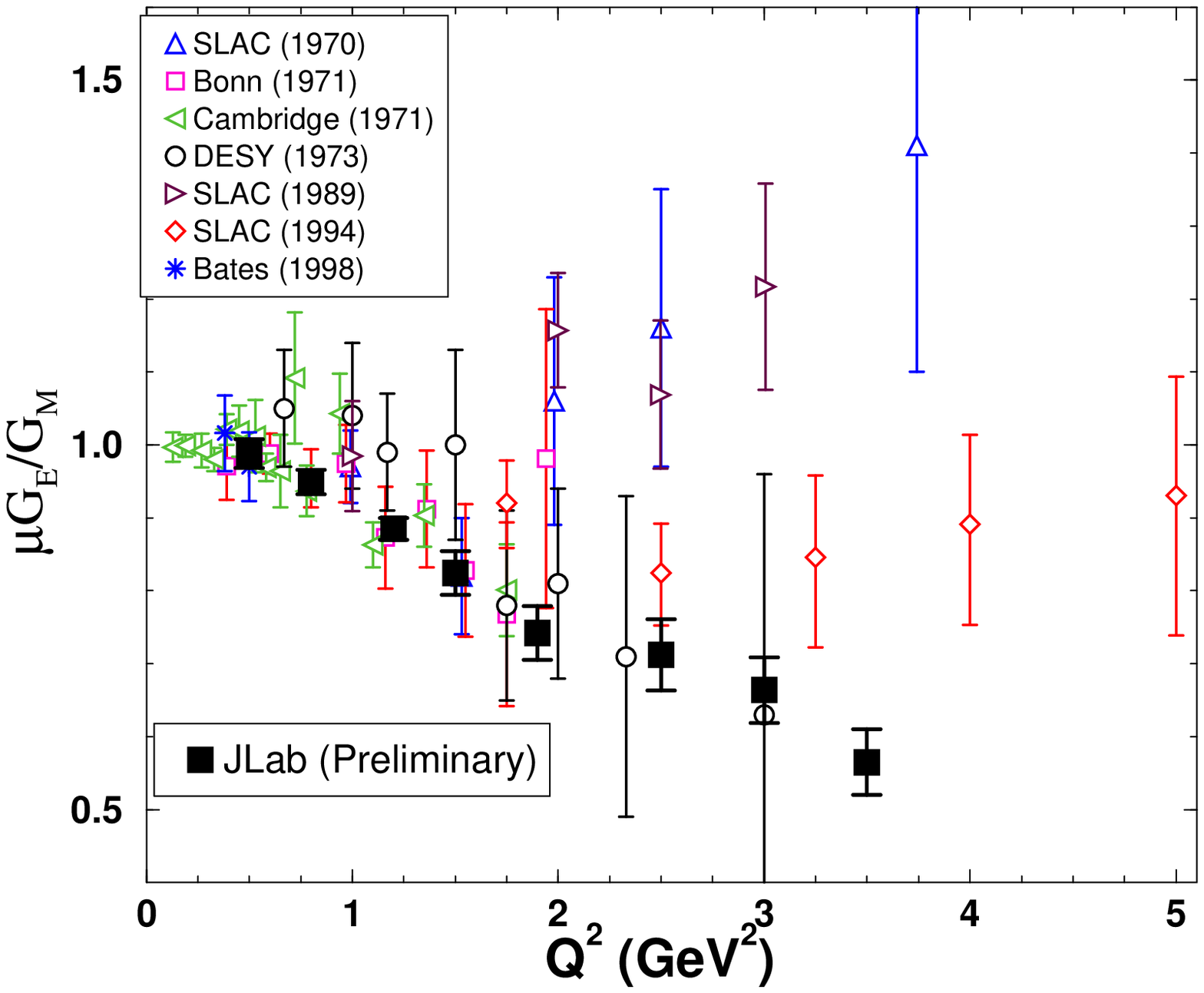}
\hsize=7.7truecm
\caption{\small Results for the ratio $R^{\gamma}_{EM}$ of electric 
and magnetic form factors of the 
proton. The full squares are preliminary results from JLAB obtained with the 
double polarization techniques \cite{perdrisat} }
\end{minipage}
\begin{minipage}{0.495\textwidth}
\epsfysize=8.5cm
\epsfbox{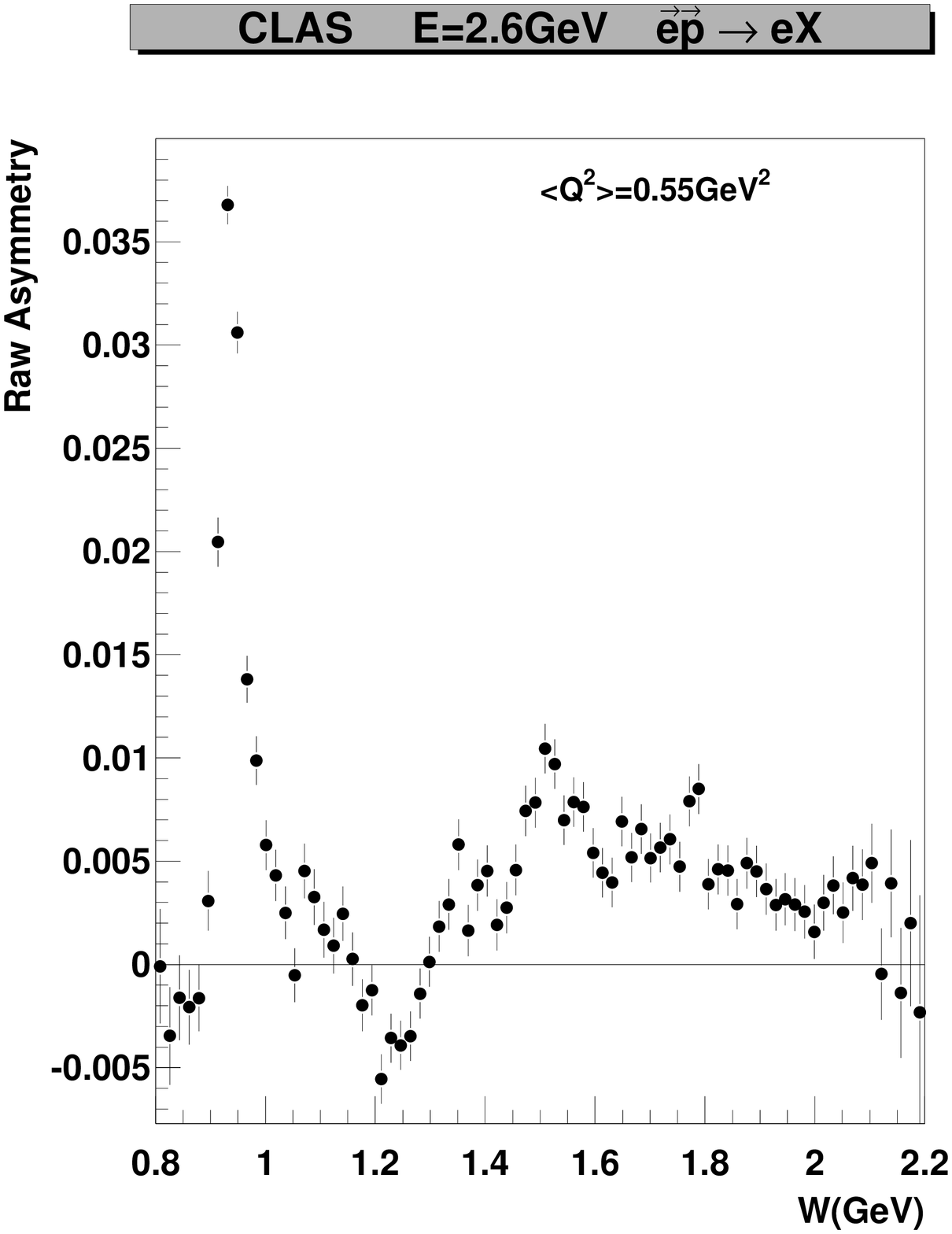}
\hsize=7.5truecm
\caption{\small Raw asymmetry measured in inclusive $\vec e N{\vec H}_3 \rightarrow e X$
scattering at JLAB.}
\end{minipage}
\end{figure}
\hsize=16.0truecm

\subsection{\bf Strangeness form factors}

From the analysis of deep inelastic polarized structure function experiments we know 
that the strange quark sea is polarized, 
and  contributes at the 5 - 10\% level to the nucleon spin.  
Then one may ask what are the strange quark contributions to the nucleon 
wave function and their corresponding form factors?
The flavor-neutral photon coupling does not distinguish s-quarks from u- 
or d-quarks. However, the tiny contribution of the $Z^o$ is parity violating, 
and allows measurement of the strangeness contribution. 
The effect is measurable due to the interference with the 
single photon graph. The asymmetry $$A_{\vec ep} = {G_FQ^2\over \sqrt{2}\pi\alpha}
{{\epsilon G_E^{\gamma} G_E^Z + \tau G_M^{\gamma}G_M^Z} \over 
{\epsilon (G_E^{\gamma})^2 + \tau (G_M^{\gamma})^2}} $$ 
\noindent
in polarized electron scattering contains 
combinations of electromagnetic and weak form factors which can be 
expressed in terms of the electromagnetic and the strangeness form factors ($G^s$). 
For example, the weak electric form factor can be written:
$$ G_E^Z = ({1\over 4} - sin^2\theta_W)G^{\gamma}_{Ep} - {1\over 4} (G^{\gamma}_{En} + G^s_E)$$
\noindent
A similar relation holds for the magnetic form factors. 
The $G^s$ form factors can be measured since the $G^{\gamma}$ form factors are known.
The elastic $\vec e p$ results of the JLAB HAPPEX experiment measured at
$Q^2 = 0.47 GeV^2$ show that strangeness contributions are small, consistent with zero, 
when measured in a combination of $G^s_E$ and $G^s_M$ \cite{happex1}:

$$ G^s_E + 0.4G^s_M = 0.023 \pm 0.034 (stat) \pm 0.022 (syst) \pm 0.026 (G^n_E)$$

At least a factor 2 smaller statistical error will 
be obtained in the 1999 run, so that the systematic error is limited by our knowledge 
of the neutron electric form factor! New measurements of $G^n_E$  
should remedy this situation \cite{day,madey}.

\section{NUCLEON SPIN STRUCTURE - FROM SHORT TO LARGE DISTANCES} 

The nucleon spin has been of central interest ever since the EMC experiment 
found that at small distances the quarks carry only a fraction of the
nucleon spin. 
Going from shorter to larger distances the quarks are dressed with gluons and $q\bar q$ pairs 
and acquire more 
and more of the nucleon spin. How is this process evolving with the distance scale? 
At the two extreme kinematic regions we have two fundamental sum rules: the 
Bjorken sum rule (Bj-SR) which holds for the proton-neutron difference in the asymptotic 
limit, and the Gerasimov Drell-Hearn sum rule (GDH-SR)at $Q^2=0$: 
$$I_{GDH} = {M^2\over 8\pi^2\alpha}\int {{\sigma_{1/2}(\nu,Q^2=0)-\sigma_{3/2}(\nu,Q^2=0)}\over \nu}d\nu = 
-{1\over 4}\kappa^2 ~~ .$$
The integral is taken over the entire inelastic energy regime. The quantity $\kappa$ is the anomalous 
magnetic moment of the target.

One connection between these regions is given by the constraint due to the GDH-SR 
- it defines 
the slope of the Bjorken integral ($\Gamma_1^{pn}(Q^2) = \int g_1^{pn}(x,Q^2)dx$) at $Q^2=0$: 
$$ I^{pn}_{GDH}(Q^2 \rightarrow 0) = 2{M^2\over Q^2}  \Gamma_1^{pn} (Q^2 \rightarrow 0) $$ 
\noindent 
Phenomenological models have been proposed to extend the GDH-SR integral for the 
proton and neutron to finite $Q^2$ 
and connect it to the deep inelastic regime \cite{buriof}. 
An interesting question is whether the transition from 
the Bj-SR to the 
GDH-SR for the proton-neutron difference can be described in terms of fundamental 
theory. 
 While for the proton and neutron alone, the GDH-SR is nearly saturated by low-lying 
 resonances \cite{burkert1,ma} with
the largest contributions coming
from the excitation of the $\Delta(1232)$, this contribution is absent in  
the pn difference. Other resonance contributions are reduced as well and 
the $Q^2$ evolution may take on a smooth transition to the Bj-SR regime. 
A crucial question in this connection is how low in $Q^2$ the Bj-SR can be evolved 
using the modern techniques of higher order QCD expansion?
Recent estimates  \cite{ji1} suggest as low as $Q^2 = 0.5$ GeV$^2$. 
At the other end, at $Q^2=0$, where hadrons are the relevant degrees of freedom,
chiral perturbation theory is applicable at very small $Q^2$, 
and may allow evolution of the GDH-SR to finite $Q^2$. Theoretical effort is 
needed to bridge the remaining gap.

{\it The importance of such efforts cannot be overemphasized as it would mark 
the first time that 
hadronic structure is described by fundamental theory in the entire kinematics
regime, from short to large distances.}

\vspace{0.2truecm}

\noindent
Experiments have been carried out at JLAB on $NH_3$, $ND_3$, and $^3He$ 
targets to extract the $Q^2$ evolution of the GDH integral for protons and neutrons
in a range of $Q^2 = 0.1 - 2.0~GeV^2$ and from the elastic to the deep 
inelastic regime. Results are expected in the year 2000. Figure 2 shows a raw asymmetry 
from an experiment on polarized $NH_3$. The positive elastic asymmetry, the negative 
asymmetry in the $\Delta$ 
region, and the switch back to positive asymmetry for higher mass resonances and 
the high energy continuum are evident.

\section{EXCITATION OF NUCLEON RESONANCES}

A large effort is being extended to the study of excited states of the nucleon. 
The  transition form factors contain information 
on the spin structure of the transition and the wave function of the 
excited state. We test predictions of baryon structure models and strong interaction 
QCD.
Another aspect is the search for, so far, unobserved states which 
are missing from the spectrum but are predicted 
by the QCD inspired quark model \cite{isgur2}. Also, are there other than $|Q^3>$ states? 
Gluonic excitations of the nucleon, i.e. $|Q^3G>$ states
should be copious, and some resonances may be ``molecules'' of baryons and mesons 
$|Q^3Q\bar Q>$.
Finding at least some of these states is important to clarify the intrinsic 
quark-gluon structure of baryons and 
the role played by the glue and mesons in hadron spectroscopy and structure. Electroproduction
is an important tool in these studies as it probes the internal 
structure of hadronic systems.

The scope of the $N^*$ program\cite{nstar,ripani} at JLAB is to 
measure many of the possible decay channels of resonances in a large kinematics
range.

\subsection{\bf The $\gamma N \Delta$ transition.}

The lowest excitation of the nucleon is the $\Delta(1232)$ ground state. The 
electromagnetic excitation is due dominantly to a quark spin flip corresponding
to a magnetic dipole transition. 
The interest today is in measuring the small electric and scalar quadrupole
transitions which are pedicted to be sensitive to possible deformation of the nucleon 
or the $\Delta(1232)$ \cite{delta}.
Contributions at the few \% level may come from the pion cloud at 
large distances, and gluon 
exchange at small distances. 
An intriguing prediction is that in the hard scattering limit the 
electric quadrupole contribution should be equal in strength to the 
magnetic dipole contribution.
 
An experiment at JLAB Hall C measured  $p\pi^o$ production in the 
$\Delta(1232)$ 
region at high momentum transfer, and found values for
$|E_{1+}/M_{1+}| < 5 \%$ at $Q^2 = 4 GeV^2$. There are no indications 
that the asymptotic value of +1 may be reached soon.

\subsection {\bf Higher mass resonances}

The inclusive spectrum shows only 3 or 4 enhancements, however 
more than 20 states are known in the mass region up to 2 GeV. 
 By measuring the electromagnetic transition of many of these 
 states we can study symmetry properties 
 between excited states and obtain a 
 more complete picture of the nucleon structure. 
For example, in the single-quark-transition model only one quark
participates in the interaction. It predicts transition amplitudes 
for a large number of states based on a few measured amplitudes.

The goal of the N* program at JLAB with the CLAS detector is to 
provide data in the entire resonance region, by measuring most 
channels, with large statistics, including many polarization observables. 
The yields of several channels recorded simultaneously are shown in Figures 3
and 4. Resonance excitations seem to be present in all channels.

These yields illustrate how the various channels have different sensitivity 
to various resonance excitations. For example, the $\Delta^{++}\pi^-$ channel 
clearly shows resonance excitation near 1720 MeV while single pion 
production is more sensitive to a resonance near 1680 MeV 
\cite{ripani}.

\begin{figure}[htbp]
\begin{minipage}{0.495\textwidth}
\epsfysize=10.0truecm
\epsfbox{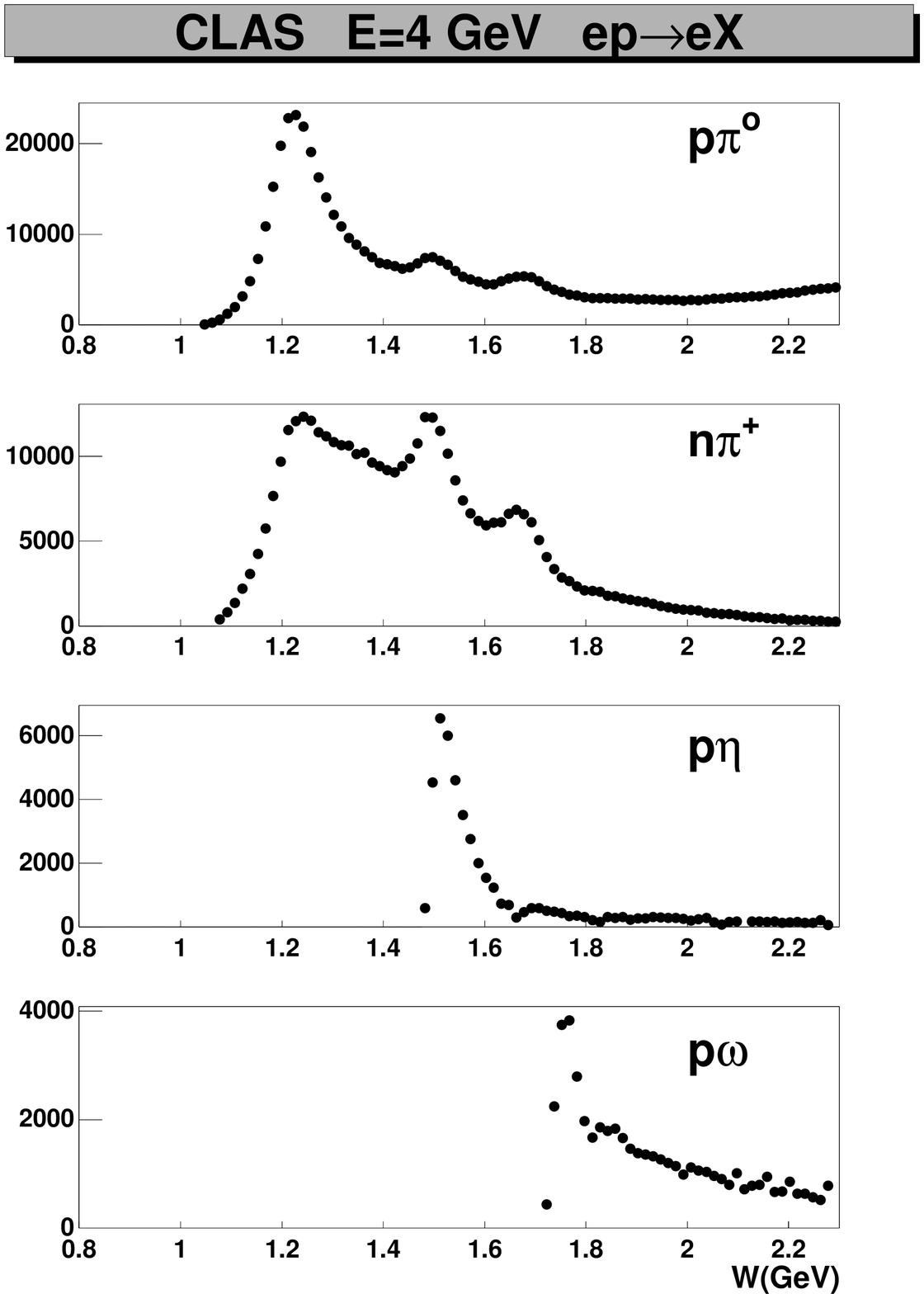}
\hsize=7.5truecm
\caption{\small Yields for various channels measured with CLAS at JLAB. The error bars are 
smaller than the data points.}
\end{minipage}
\begin{minipage}{0.495\textwidth}
\epsfysize=9.5truecm
\epsfbox{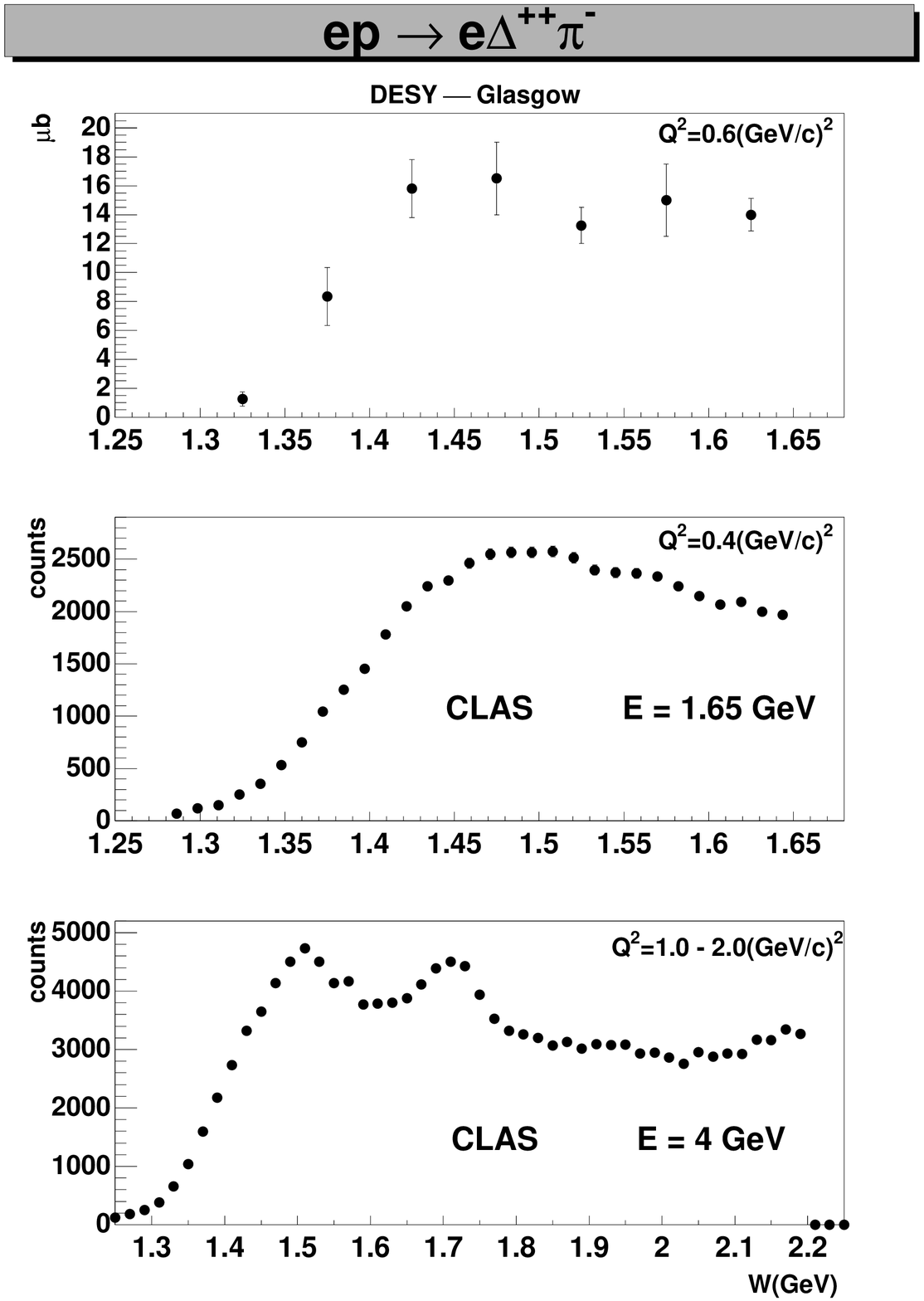}
\hsize=7.5truecm
\caption{\small Yields for the channel $\Delta^{++}\pi^-$ measured with CLAS at 
different $Q^2$ compared to previous data from DESY}
\end{minipage}
\end{figure}
\hsize=16.0truecm

\noindent The $p\omega$ channel shows resonance excitation near threshold, 
similar to the $p\eta$ channel. No resonance has been observed in this channel 
so far. For the first time  $n\pi^+$ electroproduction has been measured 
throughout the resonance region. 

Figure 4 illustrates  the vast improvement in data volume for the 
$\Delta^{++}\pi^-$ channel. 
The top panel shows DESY data taken more than 20 years ago.  The other 
panel show samples of the data taken so far with CLAS. At higher $Q^2$ 
resonance structures, not seen before in this channel are revealed.

\subsection {\bf Missing quark model states} 

These are states predicted in the $|Q^3>$ model to populate the 
mass region around 2 GeV.  However, they have
not been seen in $\pi N$ elastic scattering, our main source of information 
on the nucleon excitation spectrum. 

How do we search for these states?
Channels which are predicted to couple strongly to these states are
$N(\rho, \omega)$ or $\Delta\pi$. 
Some may also couple to $KY$ or $p\eta^{\prime}$ \cite{capstick}.

\begin{figure}[htbp]
\begin{minipage}{0.495\textwidth}
\epsfysize=9.9truecm
\epsfbox{fig_5.epsi}
\hsize=7.0truecm
\caption{\small  Electroproduction of $\omega$ mesons for different W bins. The deviation 
of the $\cos\theta$ -distribution from a smooth fall-off for the low W bin suggests significant 
s-channel resonance production.}
\end{minipage}
\begin{minipage}{0.495\textwidth}
\epsfysize=10.5truecm
\epsfbox{fig_6.epsi}
\hsize=7.5truecm
\caption{\small Ratio of resonance excitations as observed and predicted from 
deep inelastic processes using quark-hadron duality.\cite{keppel}}
\end{minipage}
\end{figure}
\hsize=16.0truecm

Figure 5 shows preliminary data from CLAS in $\omega$ production 
on protons. The 
process is expected to be dominated by diffraction-like $\pi^o$ 
exchange with strong peaking at forward $\omega$ angles, or low t,  
and a monotonic fall-off at large t. 
The data show clear deviations 
from the smooth fall-off for the W range near 1.9 GeV, were some of the 
``missing'' resonances are predicted, in comparison with the high W region. 
\cite{manak}
Although indications for resonance production are strong, analysis of 
more data and a full partial wave study are needed before 
definite conclusions may be drawn.

The  SAPHIR experiment with an analysis of just 250 $p\eta^{\prime}$ events 
at ELSA found evidence for two states with masses of 1.9 and 2.0 GeV \cite{saphir1}.
The quark model predicts indeed two resonances in this mass range 
with coupling to the $N\eta^{\prime}$ channel. 

CLAS has already collected ~50,000 $\eta^{\prime}$ events
in photo production, and 
a lot more are forthcoming later this year. Production of $\eta^{\prime}$ has 
also been observed in electron scattering for the first time with CLAS. 
This channel may also provide a new tool in the search for missing states.

$K\Lambda$ or $K\Sigma$ production may yet be another source of information on resonant
 states.  
The K$\Lambda$ data from SAPHIR \cite{saphir2} show a bump near W = 1.72 GeV, which could 
be due to resonance decay of the $P_{11}(1710)$ and $S_{11}(1650)$, both of
which couple to the K$\Lambda$ channel.
Possible resonance excitation is also seen in $K\Lambda(1520)$ production at SAPHIR,
compatible with a predicted state with a mass near 2 GeV.
New data with much higher statistics are being accumulated with the CLAS
detector, both in photo - and electro production. \cite{mestayer,schumacher} 
Strangeness production could open up a new window for light quark baryon spectroscopy,
 not available in the past.

\section{\bf QUARK-HADRON DUALITY}

I began my talk by expressing the expectation that we may eventually 
arrive at a unified description of hadronic structure from short to 
large distances. Then there should be obvious 
connections visible in the data between these regimes. 
Strong connections have indeed been observed by 
Bloom and Gilman \cite{blogil}, in the observation that the scaling curves from 
the deep inelastic cross sections also describe the average 
inclusive cross sections in the resonance region. 

This observation has recently been filled with more empirical evidence 
using inclusive ep scattering data from JLAB \cite{keppel}. 
Remarkably, elastic form factors or resonance excitations of the nucleon
 can be predicted approximately just using data from inclusive deep 
 inelastic scattering at completely different momentum transfers.  
 Figure 6 shows the ratio of measured integrals over resonance 
 regions, and predictions using deep inelastic data only. The agreement
  is surprisingly good, though not perfect, indicating that the concept 
  of duality likely is a non-trivial consequence of the underlying dynamics.

\section{QUTLOOK}

The ongoing experimental effort will provide us with a wealth of 
data in the first decade of the next millennium to address many open problems 
in hadronic structure at intermediate distances. 
{\it The experimental effort must be accompanied by a significant theoretical effort 
to translate this into real progress in our understanding of the
complex regime of strong interaction physics}.
New instrumentation will become available, e.g. the $G^o$ experiment 
at JLAB ,allowing a broad program in parity violation to 
study strangeness form factors in electron scattering in a large kinmatics range.
    
Moreover, there are new physics opportunities on the horizon. Recently, it was 
shown\cite{ji,radyu} that in exclusive processes 
the soft part and the hard part factorize for longitudinal photons 
at sufficiently high $Q^2$. 
A  new set of  ``skewed parton distributions'' can then be measured which are 
generalizations of the inclusive structure functions measured in
 deep inelastic scattering. 
For example, low-t $\rho$ production probes the unpolarized parton 
distributions, while pion production probes the polarized structure functions. 
Experiments to study these new parton distributions need to have
sufficient energy transfer and momentum transfer to reach the pQCD regime, 
high luminosity to measure the small exclusive cross sections, and
good resolution to isolate exclusive reactions.

This new area of research may become a new frontier of electromagnetic 
physics well into the next century. 

To accommodate new physics requirements, an energy upgrade in the 10-12 GeV 
range has 
been proposed for the CEBAF machine at JLAB. 
This upgrade will be accompanied by the construction of a new experimental 
hall for tagged photon experiments with a 4$\pi$ solenoid detector to study 
exotic meson spectroscopy, and production of other heavy mesons.
Existing spectrometers in Hall C will be upgraded to reach higher momenta 
and improvements of CLAS will allow it to cope with higher multiplicities.

This will give us access to kinematics where copious hybrid meson 
production is expected, higher momentum transfer 
can be reached for form factor measurements, and we may begin to map out 
the new generalized parton distributions.

\end{document}